\documentclass[conference]{IEEEtran}
\IEEEoverridecommandlockouts

\usepackage[utf8]{inputenc}
\usepackage{cite}
\usepackage{amsmath,amssymb,amsfonts}
\usepackage{algorithmic}
\usepackage{graphicx}
\usepackage{textcomp}
\usepackage[caption=false]{subfig}
\usepackage{xcolor}
\usepackage{flushend}
\def\BibTeX{{\rm B\kern-.05em{\sc i\kern-.025em b}\kern-.08em
    T\kern-.1667em\lower.7ex\hbox{E}\kern-.125emX}}

\definecolor{red}{RGB}{112,64,90}
\definecolor{green}{RGB}{90, 112, 64}    
\definecolor{blue}{RGB}{64, 90, 112}

\usepackage[bookmarks=false]{hyperref}

\hypersetup{
    pdftitle={The Archive and Package (arcp) URI scheme},    
    pdfauthor={Stian Soiland-Reyes, Marcos Cáceres},     
    pdfsubject={https://tools.ietf.org/id/draft-soilandreyes-arcp-03.html},   
    pdfcreator={Stian Soiland-Reyes},   
    pdfproducer={IEEE}, 
    pdfkeywords={Uniform resource locators, Semantic Web, Persistent identifiers, Identity management systems, Data compression, Hypertext systems, Distributed information systems, Content-based retrieval}, 
    colorlinks=true,       
    linkcolor=red,          
    citecolor=green,        
    filecolor=blue,      
    urlcolor=blue           
}

\begin{document}

\title{The Archive and Package (arcp) URI scheme
\thanks{© 2018 IEEE.
This work has been done as part of the 
\href{https://www.bioexcel.eu/}{BioExcel CoE}, a project funded by the European Commision 
(\href{https://cordis.europa.eu/projects/675728}{H2020-EINFRA-2015-1-675728}).}
}

\author{\IEEEauthorblockN{1\textsuperscript{st} Stian Soiland-Reyes}
\IEEEauthorblockA{\textit{School of Computer Science} \\
\textit{The University of Manchester} \\
Manchester, UK \\
\url{https://orcid.org/0000-0001-9842-9718}}
\and
\IEEEauthorblockN{2\textsuperscript{nd} Marcos Cáceres}
\IEEEauthorblockA{\textit{Mozilla Corporation} \\
Melbourne, Australia \\
\url{https://marcosc.com/}}
}

\maketitle

\begin{abstract}
The arcp URI scheme is introduced for location-independent identifiers to consume or reference hypermedia and linked data resources bundled inside a file archive, as well as to resolve archived resources within programmatic frameworks for Research Objects.
The Research Object for this article is available at \url{http://s11.no/2018/arcp.html#ro}
\end{abstract}

\begin{IEEEkeywords}
Uniform resource locators,
Semantic Web,
Persistent identifiers,
Identity management systems,
Data compression,
Hypertext systems,
Distributed information systems,
Content-based retrieval
\end{IEEEkeywords}

\section{Background}

Archive formats like \href{https://tools.ietf.org/html/draft-kunze-bagit-16}{BagIt} \cite{b1} have been recognized as important for preservation and transferring of datasets and other digital resources \cite{b2}. More specific examples include \href{http://co.mbine.org/documents/archive}{COMBINE} archives \cite{b3} for systems biology, \href{https://cdf.gsfc.nasa.gov/}{CDF} \cite{b4} for astronomy data, as well as the more general \href{https://support.hdfgroup.org/HDF5/doc/H5.format.html}{HDF5} \cite{b5} which is also used for meteorological data. For the purpose of this article an \textit{archive} is a collection of data files with related metadata, typically packaged in a compressed file format like \textit{.zip} or \textit{.tar.gz}.

One challenge with regards to embedding \href{https://www.w3.org/standards/semanticweb/data}{Linked Data} in such archives is how to reliably generate and resolve internal URLs, for instance \verb|<dataset13.zip>| may contain an \href{https://www.w3.org/TR/turtle/}{RDF Turtle} file \verb|<metadata/description.ttl>| to describe the CSV file \verb|<data/survey.csv>| — but in order to correctly reference that file it will either have to use a relative path \verb|<../data/survey.csv>| or some pre-existing Web URL like {\small \verb|<http://example.com/dataset13/survey.csv>|}.

The \textit{Research Object Bundle} \cite{b6} format \href{https://w3id.org/bundle/2014-11-05/}{suggested} re-using the app URI scheme for minting absolute URIs from relative paths of resources within a ZIP file. The \textit{\href{http://www.w3.org/TR/2015/NOTE-app-uri-20150723/}{app URL scheme}} \cite{b7} was originally intended for packaged web applications, where each application would get their own namespace like 
{\small \verb|<app://c6179148-3cde-4435-8e66-304453f89d59/>|} with paths resolved from the corresponding application package ZIP file. However the app URL scheme did not progress further on the W3C Recommendation track, and this approach was abandoned in favour of the combination of 
\href{https://www.w3.org/TR/appmanifest/}{Web App Manifest} \cite{b8} and \href{https://www.w3.org/TR/service-workers-1/}{Service Workers} \cite{b9}. Together these technologies reuse the http/https origin URL of a downloaded application manifest together with relative links, while also allowing a web application to work offline.

\section{The Archive and Package (arcp) URI scheme}

Inspired by the app URL scheme we defined the \href{https://tools.ietf.org/id/draft-soilandreyes-arcp-03.html}{Archive and Package (arcp) URI scheme} \cite{b10}, an IETF Internet-Draft which specifies how to mint URIs to reference resources within any archive or package, independent of archive format or location.

The primary use case for arcp is for consuming applications, which may receive an archive through various ways, like file upload from a web browser or by reference to a dataset in a repository like \href{https://zenodo.org/}{Zenodo} or \href{https://figshare.com/}{FigShare}. In order to parse Linked Data resources (say to expose them for \href{https://www.w3.org/TR/sparql11-overview/}{SPARQL} queries), they will need to generate a base URL for the root of the archive.

It should be clear that using local file URIs  \cite{b10} for extracted archives like \verb|<file:///tmp/tmp.cUK6ERfdBe/>| do not serve well for this purpose, as they are not universally unique, are difficult to create consistently, and may introduce security risks of attacks like \verb|<../../etc/passwd>|. Similarly it may be inappropriate to mint new web based URIs like \verb|<http://repo.example.com/cUK6ERfdBe/>| as web presence should not be a requirement to process a linked data archive, in particular as processing may occur on a laptop or a cloud node with no public IP address.

\subsection{Identifier structure}

By definition an arcp identifier is an URI \cite{b12} with \href{https://tools.ietf.org/id/draft-soilandreyes-arcp-03.html#rfc.section.3}{three parts}, as shown in figure \ref{fig1}.
\begin{figure}[htbp]

\centerline{\texttt{<arcp://\textcolor{red}{prefix},\textcolor{green}{namespace}\textcolor{blue}{/path}>}}
\caption{Structure of arcp identifier}
\label{fig1}
\end{figure}

The arcp Internet-Draft specifies three initial \texttt{\textcolor{red}{prefix}} values: \verb|uuid|, \verb|ni| and \verb|name|, each which defines how to identify a particular archive by a corresponding \texttt{\textcolor{green}{namespace}}. These namespaces are not intended to be directly resolvable without prior knowledge of the corresponding archive.

The \texttt{\textcolor{blue}{path}} is the folder and file path within the archive, represented as an 
\href{https://tools.ietf.org/html/rfc3986#section-3.3}{URI path} \cite{b12} e.g. \texttt{/file.txt} or 
\texttt{/my\%20project/about/intro.doc} — using percent-escaping if needed. The root folder \texttt{/} represent the archive itself.

\subsection{UUID-based identifiers}

The simplest case for temporary \href{https://tools.ietf.org/id/draft-soilandreyes-arcp-03.html#rfc.appendix.A.1}{sandbox} processing of an archive with arcp is to generate a new random \href{https://tools.ietf.org/html/rfc4122#section-4.4}{UUIDv4} \cite{b13}, e.g.:

{\footnotesize 
\begin{verbatim}
c6179148-3cde-4435-8e66-304453f89d59
\end{verbatim}
}

From this the corresponding arcp URI is:

{\footnotesize 
\begin{verbatim}
<arcp://uuid,c6179148-3cde-4435-8e66-304453f89d59/>
\end{verbatim}
}

This \textit{base URI} can be used when resolving relative URI references, e.g. if 
\texttt{<metadata/description.ttl>} references 
\texttt{<../data/survey.csv>} we find the absolute URIs:

{\footnotesize 
\begin{verbatim}
<arcp://uuid,c6179148-3cde-4435-8e66-304453f89d59
  /metadata/description.ttl>
<arcp://uuid,c6179148-3cde-4435-8e66-304453f89d59
  /data/survey.csv>
\end{verbatim}
}

The application is then able to do translation from arcp to local paths using URI parsing libraries to select the URI path, and augment that to the locally extracted path. Such arcp identifiers are temporary in nature, but the application can maintain a mapping from the UUID to the archive and perform extraction on demand, or the archive can \href{https://tools.ietf.org/id/draft-soilandreyes-arcp-03.html#rfc.appendix.A.4}{self-declare} its UUID, such as the \verb|External-Identifier| header in BagIt \cite{b1}.

arcp also suggests how a UUID can be reliably created from the URL \href{https://tools.ietf.org/id/draft-soilandreyes-arcp-03.html#rfc.appendix.A.2}{location} of an archive. For instance, an application may be processing a file from:

{\footnotesize 
\begin{verbatim}
http://example.com/download/archive13.zip>
\end{verbatim}
}

The application can calculate the \href{https://tools.ietf.org/html/rfc4122#section-4.3}{\textit{name-based UUIDv5}} \cite{b13}
by SHA1 hashing the URL string and mint:

{\footnotesize 
\begin{verbatim}
<arcp://d9f0b57d-0504-5e9a-abae-f5f2b8c49b94/>
\end{verbatim}
}

With this method anyone processing that archive URL will always get the same arcp base URI, however the application will still need to maintain a mapping to find the original archive URL. Location-based arcp identifiers may also not be ideal for preservation purposes, as the archive might change upstream or move to a different location.

\subsection{Hash-based identifiers}

For this arcp defines a \href{https://tools.ietf.org/id/draft-soilandreyes-arcp-03.html#rfc.appendix.A.3}{hash-based method}, where the bytes of the archive file is used to find a checksum-based identifier based on the 
\href{https://tools.ietf.org/html/rfc6920}{Naming Things With Hashes} (ni) URI scheme  \cite{b14}. For instance if the sha-256 checksum of a Zip file is in hexadecimal:

{\footnotesize 
\begin{verbatim}
7f83b1657ff1fc53b92dc18148a1d65d
fc2d4b1fa3d677284addd200126d9069
\end{verbatim}
}

After base64 encoding the \texttt{ni:} uri would be:

{\footnotesize 
\begin{verbatim}
<ni:///sha-256;
  f4OxZX_x_FO5LcGBSKHWXfwtSx-j1ncoSt3SABJtkGk>
\end{verbatim}
}

The corresponding arcp base URIs for resources within the archive is thus:

{\footnotesize 
\begin{verbatim}
<arcp://ni,sha-256;
  f4OxZX_x_FO5LcGBSKHWXfwtSx-j1ncoSt3SABJtkGk/>
\end{verbatim}
}

With this method, anyone processing the byte-wise equal archive (using the same hash method) will get the same identifier. 

Another advantage is that hash-identified archives can be retrieved from a \href{https://tools.ietf.org/html/rfc6920#section-4}{NI resolver} \cite{b14} using well known paths \cite{b15}:

{\footnotesize 
\begin{verbatim}
<http://repo.example.com/.well-known/ni/sha-256
  /f4OxZX_x_FO5LcGBSKHWXfwtSx-j1ncoSt3SABJtkGk>
\end{verbatim}
}

Clients can verify the checksum of the downloaded archive, so any accepting resolver endpoint can be used.

\subsection{Name-based identifiers}

Finally, paying homage to its origin in app URLs, arcp can use a system-based app name. This is a suggested mechanism for resolving resources of an application package installed in a runtime system like \href{https://developer.android.com/studio/build/application-id}{Android applicationId} or Java package name, where an application identifier can be directly reused in arcp for URIs within that runtime system, e.g. to reference the resource \verb|styles/resource1.css| within the installed package \verb|com.example.myapp| one can use the URI:

{\footnotesize 
\begin{verbatim}
<arcp://name,com.example.myapp/styles/resource1.css>
\end{verbatim}
}

As application package content do not necessarily correspond to archive file listings, it is open-ended how name-based arcp identifiers can be resolved, and indeed package content may vary per operating system, device type or application version, and so name-based arcp identifiers should be treated as system-local identifiers similar to \verb|file:///| URIs  \cite{b11}, but within a particular programming framework.

\section{Related work}

\subsection{Archive fragments}

Without using arcp one could in theory still reference files within archives at an URL with \texttt{\#} fragments:

{\footnotesize 
\begin{verbatim}
<http://example.com/download
  /archive13.zip#data/survey.csv>
\end{verbatim}
}

Unlike formats like \href{https://www.iana.org/assignments/media-types/text/html}{\textit{text/html}}
or \href{https://www.iana.org/assignments/media-types/application/pdf}{\textit{application/pdf}}, most archive media formats like \href{https://www.iana.org/assignments/media-types/application/zip}{\textit{application/zip}} unfortunately do not define a fragment syntax, and some major types like \textit{tar.gz} are not even listed in the 
\href{https://www.iana.org/assignments/media-types/}{IANA media types registry}. Therefore this would be an ad-hoc approach which still needs to clarify details in order to be interoperable, for instance character escaping, if the root is \verb|#| or \verb|#/|, and how to reference nested fragment identifiers in hypermedia within archived resources.

\subsection{File URIs}

As argued above, file URLs \cite{b11} that represent local directories are fragile and not globally unique. It is perhaps less known that file URLs 
\href{https://tools.ietf.org/html/rfc8089#section-2}{can specify a host name}:

{\footnotesize 
\begin{verbatim}
<file://host.example.com
  /home/alice/extracted/archive13/>
\end{verbatim}
}

The above references a file path on the machine with the fully qualified domain name (FQDN)
\texttt{host.example.com}. The usually empty hostname is equivalent to \texttt{localhost}.

This approach may be used if both the hostname and extracted path are stable (e.g. a repository file server), but this faces the same challenges as minting http/https URLs, which in many cases would be preferable as they are also globally resolvable. 

An ad-hoc possibility here could be to use a UUID \cite{b13} as "hostname" to represent an archive's internal file system:

{\footnotesize 
\begin{verbatim}
file://8f26cb8c-617e-46b4-bc48-e650bf70f33d
  /data/survey.csv/>
\end{verbatim}
}

This is technically permittable as the \verb|file:| URL scheme \cite{b11} do not define any particular connection protocols, and an UUID is unlikely to be a valid hostname in DNS. Such \texttt{file:} URIs could however cause confusion against file paths on \texttt{localhost}, for instance Firefox 62.0 opens {\footnotesize \url{file://8cd4ce0d-4a41-4b4e-bfdd-1e2d0495f714/}} to browse the local file system.

\subsection{JAR URLs}

If we restrict usage to ZIP files at a known URL, then they are in theory also valid JAR files, and we can address files with the 
\href{https://docs.oracle.com/javase/9/docs/api/java/net/JarURLConnection.html}{\textit{jar URL scheme}}:

{\footnotesize 
\begin{verbatim}
<jar:http://example.com
  /download/archive13.zip!/data/survey.csv>
\end{verbatim}
}

Here relative URIs may not parse well, as it is easy to accidentally climb out of \verb|!/|, and technically the JAR URI scheme is missing the familiar \verb|://| to indicate for URI parser libraries that it is indeed an \href{https://tools.ietf.org/html/rfc3986#section-1.2.3}{hierarchical URI scheme}  \cite{b12}.

\subsection{Object Reuse and Exchange proxies}

\href{http://www.openarchives.org/ore/}{OAI-ORE} \cite{b16} defines \href{http://www.openarchives.org/ore/1.0/datamodel#Proxy}{proxies} to represent a resource as aggregated in a collection; these can be used to model archives  \cite{b17}, but ORE proxies face two problems: How to represent the file path, and how to identify the proxy so it can be used as a reference in Linked Data. The resource must be identified using two triples of \texttt{ore:proxyFor} (the archived file) and \verb|ore:proxyIn| (the archive); but this reduces to the same problem of identifying the file. The ni URI \cite{b14} for the file bytes can in theory be used to identify the file, but the other missing information is the file path and name, which usually convey meaning for users.

The Research Object ontology’s 
\href{https://w3id.org/ro/2016-01-28/ro#FolderEntry}{\texttt{FolderEntry}} specializes the \texttt{ore:Proxy} to add a property \texttt{ro:entryName} to indicate the filename, as exemplified in figure \ref{figORE}, but to find the full archive file path one would have to traverse the parent folder’s \texttt{ro:entryName}. In either case there is no defined method to predictably generate unique identifiers for the ORE proxies themselves, although the 
\href{https://w3id.org/bundle/2014-11-05/}{RO Bundle} specification recommend they should be randomly generated \texttt{urn:uuid} URIs, which would not be compatible with relative URIs within an archive. 

\begin{figure*}[hbtp]
{\footnotesize 
\begin{verbatim}
@prefix ore: <http://www.openarchives.org/ore/terms/> .
@prefix ro: <http://purl.org/wf4ever/ro#> .
<urn:uuid:c5971b62-72e6-4a8f-8b0b-944065e0d5c8> a ore:Proxy, ro:FolderEntry ;
    ore:proxyFor <ni:///sha-256;f4OxZX_x_FO5LcGBSKHWXfwtSx-j1ncoSt3SABJtkGk> ;
    ore:proxyIn <urn:uuid:efb14c0a-3cd5-4d78-a168-f246d18bde39> ;
    ro:entryName "survey.csv" .
<urn:uuid:efb14c0a-3cd5-4d78-a168-f246d18bde39> a ore:Aggregation, ro:Folder .
<urn:uuid:24b34ecb-e46b-46ec-be36-a18dbba90247> a ore:Proxy, ro:FolderEntry ;
    ore:proxyFor <urn:uuid:efb14c0a-3cd5-4d78-a168-f246d18bde39> ;
    ore:proxyIn <http://example.com/download/archive13.zip> ;
    ro:entryName "data/" .
\end{verbatim}
}
\caption{RDF Turtle example of how a file with the sha256 checksum \texttt{7f83b1…6d9069} could be described using RO folders and ORE proxies to belong 
to \texttt{<data/survey.csv>} within the archive downloaded from \texttt{<http://example.com/download/archive13.zip>}
}
\label{figORE}
\end{figure*}

\subsection{Publishing file systems as Linked Data}

F2R \cite{b18}, using the \href{http://oscaf.sourceforge.net/nfo.html}{Nepomuk File Ontology} \cite{b19}, defines a way to publish file systems as Linked Data, where a server endpoint exposes the files and their file system metadata. 

F2R URIs are localized to an endpoint and an free-text named file system, e.g. \texttt{mysource}, and files are identified with UUIDs: 

{\footnotesize 
\begin{verbatim}
<http://f2r.example.com
  /mysource/09b205be-bj80–4ab9–8ddc-802be95220bb>
\end{verbatim}
}

Using the same example as for OAI-ORE we can combine F2R with PAV \cite{b20}, as shown in figure \ref{figNepomuk}.

\begin{figure*}[htbp]
{\footnotesize 
\begin{verbatim}
@base <http://f2r.example.com/mysource/> .
@prefix nfo: <http://www.semanticdesktop.org/ontologies/2007/03/22/nfo#> .  
<c5971b62-72e6-4a8f-8b0b-944065e0d5c8> a nfo:ArchiveItem;
    nfo:fileName "survey.csv" ;
    nfo:belongsToContainer <24b34ecb-e46b-46ec-be36-a18dbba90247> .
<24b34ecb-e46b-46ec-be36-a18dbba90247> a nfo:ArchiveItem;
    nfo:fileName "data" ;
    nfo:belongsToContainer <5d0a538a-ef00-48b6-bcb2-f561effe9fe5> .
<5d0a538a-ef00-48b6-bcb2-f561effe9fe5> a nfo:ArchiveItem:
    nfo:fileName "archive13.zip" ;
    nfo:belongsToContainer <http://f2r.example.com/mysource/> ;
    pav:retrievedFrom <http://example.com/download/archive13.zip> .
<http://f2r.example.com/mysource/> a nfo:Filesystem .
\end{verbatim}
}
\caption{RDF Turtle description of a file \texttt{<data/survey.csv>} within an archive \texttt{<http://example.com/download/archive13.zip>}, using Nepomuk File Ontology \cite{b19}, PAV \cite{b20} and F2R \cite{b18} identifiers.}
\label{figNepomuk}
\end{figure*}

The F2R approach have similar disadvantages as JAR and OAI-ORE; in that the URIs do not support relative path resolution, that a web endpoint must be set up, and that the file paths are hidden through multiple steps. In addition one would need to assigned a corresponding file system name like \verb|mysource|, although one may use a single file system as exemplified above and use \verb|belongsToContainer| to treat archive files as if they are folders.

\subsection{EPUB canonical fragment identifiers}

\href{https://www.w3.org/Submission/epub31/}{EPUB} is a standard for hypermedia eBooks. 
\href{https://w3id.org/bundle/2014-11-05/#ucf}{RO Bundle} \cite{b6} is based on the 
\href{https://www.w3.org/Submission/2017/SUBM-epub-ocf-20170125/}{EPUB Open Container Format} \cite{b21}. 
\href{http://www.idpf.org/epub/linking/cfi/}{EPUB Canonical Fragment Identifiers} \cite{b22} can link to nested XML elements of an publication using a variation of 
\href{https://www.w3.org/TR/xpath20/}{XPath} with 
\href{https://www.idpf.org/epub/linking/cfi/epub-cfi.html#sec-path-child-ref}{doubled indexes}: 

{\footnotesize 
\begin{verbatim}
<http://example.com/book.epub
  #epubcfi(/6/4[chap01ref]!/4[body01]/10[para05])>
\end{verbatim}
}

The above example show an example to a paragraph with an ePub book. Here \texttt{/6} refer to the 3rd element of the root manifest’s 
\href{http://www.idpf.org/epub/31/spec/epub-packages.html#sec-package-elem}{\texttt{<package>}} element (which in ePub is always 
\href{http://www.idpf.org/epub/31/spec/epub-packages.html#elemdef-opf-spine}{\texttt{<spine>}}), then \texttt{/4[chap01ref]} is the second element 
\href{http://www.idpf.org/epub/31/spec/epub-packages.html#elemdef-spine-itemref}{\texttt{<itemref>}} with \verb|xml:id="chap01ref"|. 

The \texttt{!} character means the element's reference is followed to open the corresponding XML file, where \verb|/4[body01]| is the 2nd element with id \verb|body01|, traversed to find the 5th element with id \verb|para05|.

While this is quite a powerful construct that can refer to any XML element of nested documents, even sentences or words, it seems rather contrived and inflexible. The major limitation is that ePub archive resources are not identified by file paths, but must be addressable through rather rigid XML structures (order can’t change), thus this approach is not appropriate for archives without an XML manifest. Even if using a RDF/XML manifest it would be inadvisable to assume a fixed order of it’s XML elements. It seems however an appropriate reference scheme for ePub documents, which generallyhave a fixed reading order.

\section{arcp implementations}

The 
\href{http://arcp.readthedocs.io/}{\textbf{arcp Python library}} \cite{b23} was developed to help creating, parsing and validating arcp URIs. In particular it can 
\href{http://arcp.readthedocs.io/en/0.2.0/generate.html}{generate arcp} based on random UUIDs, URL locations, names and hashing archive bytes. The \href{http://arcp.readthedocs.io/en/0.2.0/parse.html}{arcp parser} recognize the arcp prefix and can extract UUIDs or hashes, and can generate the corresponding \verb|.well_known/ni| URI for retrieving the archive. This library is meant to complement the Python 3 \textit{urlparse} library, and so it is deemed out of scope for this library to do resolution of arcp based on archive or network access.

The 
\href{https://github.com/apache/incubator-taverna-language/tree/master/taverna-robundle}{\textbf{Research Object Bundle library}}, part of 
\href{https://taverna.incubator.apache.org/download/language/}{Apache Taverna (incubating)}, is 
\href{https://issues.apache.org/jira/browse/TAVERNA-1037}{adding support for arcp URIs} in its opening and creation of RO bundles, initially using the arcp UUID format as a replacement for app URIs, with planned support also for hash-based identifiers and opening RO Bundles from a \verb|.well-known/ni| endpoint.

The \href{https://w3id.org/cwl/prov}{\textbf{CWLProv}} \cite{b24} approach for capturing provenance of executing Common Workflow Language is using arcp in its BagIt metadata 
\href{https://github.com/common-workflow-language/cwlprov/blob/0.4.0/examples/revsort-run-1/bag-info.txt#L5}{\texttt{bag-info.txt}} using  \verb|External-Identifier| to identify its research object:

{\footnotesize 
\begin{verbatim}
External-Identifier: 
  arcp://uuid,d47d3d43-4830-44f0-aa32-4cda74849c63/
\end{verbatim} 
}

For CWLProv the use of arcp is crucial, as it assigns global identifiers for use across resources in the RO bag, including the 
\href{https://github.com/common-workflow-language/cwlprov/blob/0.4.0/examples/revsort-run-1/metadata/manifest.json#L4}{RO manifest itself} and in W3C PROV file formats like 
\href{https://github.com/common-workflow-language/cwlprov/blob/0.4.0/examples/revsort-run-1/metadata/provenance/primary.cwlprov.provn}{PROV-N} and 
\href{https://github.com/common-workflow-language/cwlprov/blob/0.4.0/examples/revsort-run-1/metadata/provenance/primary.cwlprov.nt}{N-Triples}, as neither format  support relative URIs.

In this approach the UUID component of the RO arcp identifier {\footnotesize  \texttt{d47d3d43-4830-44f0-aa32-4cda74849c63}} also appears in the 
\href{https://github.com/common-workflow-language/cwlprov/blob/0.4.0/examples/revsort-run-1/metadata/provenance/primary.cwlprov.provn#L23}{workflow provenance} as the identifier of the top-level workflow run (a PROV Activity):

{\footnotesize 
\begin{verbatim}
prefix id <urn:uuid:>
activity(id:d47d3d43-4830-44f0-aa32-4cda74849c63, 
  2018-08-21T17:26:24.467636, -, 
  [prov:type='wfprov:WorkflowRun', 
  prov:label="Run of workflow/packed.cwl#main"])
\end{verbatim} 
}

This is showcasing how an RO that is the primary representation of a  \textit{non-information resource} (e.g. a process) can be identified using a directly derived arcp URI. While this could in theory also been achieved with an arcp UUIDv5 derived from hashing the URI “location” of the activity, 
that would be a confusing hack, as \texttt{urn:uuid:} references by design are not resolvable, and hence technically not URLs. UUIDv5 hashing could however be appropriate for non-information resource if they have a resolvable http/https permalink.

\section{Conclusion}

This article propose the arcp identifier scheme for resources within archives using formats like ZIP, tar and BagIt, and suggest arcp is useful for identifying standalone Research Objects and for processing Linked Data embedded in archives. The Internet-Draft \verb|draft-soilandreyes-arcp| \cite{b10} is under consideration by IETF’s Applications and Real-Time Area to progress towards Informational RFC status.

\vspace{12pt}

\end{document}